\documentclass{aa}
\usepackage[varg]{txfonts}
\usepackage{graphicx}
\usepackage{xcolor}
\usepackage{hyperref}
\hypersetup{
    colorlinks=true,
    linkcolor=black,     
    urlcolor=blue,
    citecolor=blue
    }
\usepackage{ulem}

\begin{document}
\title{Spacetime in motion: an evolving relativistic binary black hole metric for GIZMO}
\titlerunning{MBHB in GIZMO}
\author{Giacomo Fedrigo$^{1,2}$, Alessandro Lupi$^{1,2,3}$, Alessia Franchini$^{2,4}$ and Matteo Bonetti$^{2,5,6}$}
\date{}

\institute{$^1$Como Lake centre for AstroPhysics, DiSAT, Università dell’Insubria, via Valleggio 11, 22100 Como, Italy \\$^2$INFN, Sezione di Milano-Bicocca, Piazza della Scienza 3, I-20126 Milano, Italy \\$^3$INAF - Osservatorio di Astrofisica e Scienza dello Spazio di Bologna, Via Gobetti 93/3, I-40129 Bologna, Italy\\$^4$Dipartimento di Fisica ``A. Pontremoli'', Università degli Studi di Milano, Via Giovanni Celoria 16, 20134 Milano, Italy
\\$^5$Dipartimento di Fisica ``G. Occhialini'', Universit\`a degli Studi di Milano-Bicocca, Piazza della Scienza 3, I-20126 Milano, Italy
\\$^6$INAF - Osservatorio Astronomico di Brera, via Brera 20, I-20121 Milano, Italy}

\abstract
    {The last evolutionary stages of massive black hole binaries prior to coalescence is dominated by the emission of gravitational waves, which will be probed by the future Laser Interferometer Space Antenna. If gas is present around the two black holes, however, the associated electromagnetic emission can provide additional information about the binary properties and location before the merger event. For this reason, a proper characterisation of the electromagnetic emission during these phases is of fundamental importance, and requires a detailed description of the gas dynamics close to the event horizon of the two black holes, only achievable via numerical simulations. Within this context, we present the implementation of the Superposed Kerr-Schild dynamic metric in the relativistic scheme in the meshless code \textsc{gizmo}.
    Our code can now simulate black hole binaries approaching merger with high computational efficiency and accuracy, taking into account relativistic effects on the gas. To validate our implementation, we perform two tests. First, we explore the case of a relativistic Bondi flow around a binary, finding very good agreement with numerical relativity simulations. Then we explore the case of an inviscid relativistic circumbinary disc, comparing our results with a similar simulation run assuming Newtonian gravity. In this second case, we find moderate differences in the mass accretion rate and in the inflow dynamics, which suggest that the presence of a non-Keplerian potential and of apsidal precession in the orbiting gas trajectories may produce stronger shocks and boost angular momentum transport in the disc. Our work highlights the importance of accounting for relativistic corrections in accretion disc simulations around black hole binaries approaching merger, even at scales much larger than those currently probed by numerical relativity simulations.}
 
\maketitle

\section{Introduction}\label{sec:Introduction}

In the next decade, the LISA mission (Laser Interferometer Space Antenna) is expected to detect gravitational wave (GW) signals in the mHz regime sourced by massive black hole binaries (MBHBs) approaching merger \citep{Kocsis2006,Dotti2006,Mayer2007,AmaroSeoane2012,Capelo2015,AmaroSeoane2017}. These sources reside in the centre of galaxies and are expected to evolve and eventually merge in gas-rich environments. The presence of gas and the highly dynamic spacetime suggest the possibility of accretion occurring on the binary components, resulting in the production of strong electromagnetic (EM) signals.
Therefore, MBHBs close to merger represent the perfect scenario for multimessenger astronomy with massive compact objects. The coincident detection of a GW signal and an EM counterpart could provide unique information about the system environment and evolution.

Furthermore, the Pulsar Timing Array (PTA) community has recently announced the detection of a stochastic gravitational wave background signal in the nHz frequency band \citep{EPTA2023, NANOGrav2023, Parkers2023, CPTA2023}, likely produced by a population of $10^8-10^9 M_\odot$ BH binaries. These PTA results demonstrated our deep understanding of the GW signal produced by massive compact objects, while giving a first proof of the existence of close-separation MBHBs.

Still, there is a huge uncertainty about the signature features of an EM signal produced by a MBHB that make the source distinguishable from a single AGN \citep{Bogdanovic2011,Dotti2012, Bogdanovic2022} and that cannot be explained by other mechanisms, e.g. asymmetric emissions or periodic modulation of an AGN disc \citep{Rigamonti2025, Dotti2023}.
Major differences may exist in the accretion flows close to the central binary. The non-axisymmetric potential of a binary could carve an empty cavity around the BHs thanks to strong torques, while thin streams of gas fall into the centre and get accreted by the compact objects \citep{MacFadyen2008, Noble2012}, yielding very characteristic EM features. The dynamics of these processes may depend strongly on the spins of the BHs and on the properties of the gaseous environment \citep{Bowen2018,Bode2010,Cattorini2021,Ruiz2023}. Also, variability in the emitted light curves is observed in many simulations \citep{Krauth2023,Gutierrez2022,Franchini2024,Cocchiararo2024,Cocchiararo2025b} and could be a smoking gun for detecting an evolving MBHB with optical facilities such as the Vera Rubin Observatory's LSST \citep{Chiesa2025}.

Numerical simulations are powerful tools to study and understand such extreme environments and can help predict the EM emission from a MBHB approaching merger. Many works aim to reach a stationary regime to study the properties of a circumbinary disc. Therefore, starting from idealized initial condition, it is necessary to simulate the system for hundreds of orbits, as discussed in the binary disc code comparison presented in \citet{Duffell2024}.
Hence, many works have focused on long-term Newtonian simulations, both in 2D and in 3D \citep{Munoz2020,Krauth2023}, also considering post-Newtonian corrections to the binary evolution \citep{Franchini2024}.\\On the other hand, many interesting accretion features exclusive to a MBHB system may reside close to the event horizons \citep{dAscoli2018,Gold2019,CattoriniGiacomazzo2024}.
To accurately explore those regions, 3D numerical relativity simulations are necessary. However, these kinds of simulations are extremely computationally expensive and are therefore limited to evolving BH binaries only for a few orbits before merger \citep{Bode2010,Giacomazzo2012,Gold2014,Paschalidis2021,Fedrigo2024}. Only recently, hybrid and approximated methods have been employed to solve this issue and evolve the system for a longer time \citep{MignonRisse2023,Avara2024,Ennoggi2025,Ennoggi2025b}.

A simpler solution has been presented by \citet{Combi2021} and \citet{Combi2024} in the form of an approximated dynamic metric, created by the superposition of two boosted BHs in Kerr-Schild coordinates, hence called Superposed Kerr-Schild (SKS) metric. It represents an accurate yet computationally cheap way to evaluate the dynamic metric of a BH binary, and it can be used to perform general relativistic magnetohydrodynamics (GRMHD) or ray-tracing simulations. Thanks to the high versatility of the SKS metric, it is now possible to deeply explore the evolution and dynamics of the gaseous environment in which MBHBs reside using different state-of-the-art astrophysical codes.

In this work, we present the implementation of the SKS metric in the relativistic meshless version of the \textsc{gizmo} code \citep{GizmoHD,Lupi2023,Fedrigo2025}. Our goal is to build a code that can evolve relativistic circumbinary discs around BH binaries for thousands of orbits and to study their properties in search of unique signatures. 

In Section~\ref{sec:SKSFormulation} we quickly review the SKS formulation and in Sec \ref{sec:SKSImplementation} we explain some details about its implementation in \textsc{gizmo}. Then, in Section~\ref{sec:Tests} we report results from two tests we performed to check the correct implementation of the metric and its stability within the code. The first one is a comparison between our results and a numerical relativity simulation of a uniform gas distribution accreting onto a BH binary (Section~\ref{sec:Bondi}). With the second test, we explore the differences between a Newtonian and a relativistic circumbinary disc (Section~\ref{sec:CBD}). Finally, we draw our conclusions in Section~\ref{Sec:Conclusion}.

\section{Black hole binary metric}\label{sec:2}
In this section, we describe the implementation of the SKS metric in \textsc{gizmo}. The relativistic scheme implemented in \textsc{gizmo} is presented in \citet[][hydrodynamics]{Lupi2023} and \citet[][magnetohydrodynamics]{Fedrigo2025}. Note that we do not consider magnetic fields in the tests discussed in this work, deferring the study of magnetised plasma in a MBHB environment to a future study.
The superposed binary metric was first introduced in \citet{Armengol2021}, later refined in \citet{Combi2021}, and finally extended to comprehend the merger phase in \citet{Combi2024} [hereafter Co24]. Here, we briefly review the main concepts of the Co24 metric formulation, and we then focus on some implementation details. Throughout this work, we employ natural units, i.e. $G=c=1$.

\subsection{SKS formulation}\label{sec:SKSFormulation}

The metric of a single spinning BH of mass $M$ and spin $\textbf{a}$ can be expressed in the ``Cartesian'' Kerr-Schild (KS) coordinate system $\{X^\mu\}$:
\begin{equation}\label{eq:KerrSchildMetric}
    g_{\mu\nu} = \eta_{\mu\nu} + 2\mathcal{H}l_\mu l_\nu,
\end{equation}
where the null vector $l$ is defined by
\begin{equation}\label{eq:KerrSchildLaLb}
    l_\mu dX^\mu= dt + \frac{1}{r^2+a^2}\left[rX^i - \epsilon^i_{jk}a^jX^k  + \frac{a^iX^j\delta_{ij}}{r}a^i\right]dX_i,
\end{equation}
the function $\mathcal{H}$ is
\begin{equation}
    \mathcal{H}(X^i) = \frac{Mr^3}{r^4+(a^iX^j\delta_{ij})^2},
\end{equation}
the Boyer-Lindquist radius is 
\begin{equation}\label{eq:KerrSchildRadius}
    r=\sqrt{\frac{R^2-a^2+\sqrt{(R^2-a^2)^2+4(a^iX^j\delta_{ij})^2}}{2}},
\end{equation}
and $R^2 = X^2+Y^2+Z^2$.\\

Following Co24, a generalised Lorentz boost transformation is then applied to the KS metric (\ref{eq:KerrSchildMetric}), going from the non-inertial frame at rest with the BH $\{X^\mu\}$ to an inertial frame where the BH is moving on an inspiraling orbit $\{x^\mu\}$. Defining the BH trajectory with $\textbf{s}(t)$, the coordinate transformation with velocity $\textbf{v} = \beta\textbf{n}(t)$ is given by:
\begin{equation}\label{eq:LorentzBoost}
    \begin{split}
        &T = t-W\beta(n_x\bar{x} + n_y\bar{y}+n_z\bar{z}),\\
        &X = \bar{x}[1+n_x^2(W-1)] + \bar{y}(W-1)n_xn_y + \bar{z}(W-1)n_xn_z,\\
        &Y = \bar{y}[1+n_y^2(W-1)] + \bar{x}(W-1)n_xn_y + \bar{z}(W-1)n_yn_z,\\
        &Z = \bar{z}[1+n_z^2(W-1)] + \bar{y}(W-1)n_yn_z + \bar{x}(W-1)n_xn_z,\\
    \end{split}
\end{equation}
where $\bar{x}^i = x^i - s^i(t)$ and $W$ is the Lorentz factor.
To conclude, the superposed metric of two inspiralling spinning BHs in KS coordinates can be written as
\begin{equation}\label{eq:SKSmetric}
    g_{\mu\nu} = \eta_{\mu\nu} + \left[2\frac{\partial X^\rho}{\partial x^\mu}\frac{\partial X^\sigma}{\partial x^\nu}\mathcal{H}l_\rho l_\sigma\right]_1 + \left[2\frac{\partial X^\rho}{\partial x^\mu}\frac{\partial X^\sigma}{\partial x^\nu}\mathcal{H}l_\rho l_\sigma\right]_2.
\end{equation}

This approximate metric formulation exhibits good accuracy and low constraint violation for intermediate binary separations, but it begins to fail at BH separations $\lesssim 8 M$. The SKS metric has good performances and is smooth at the ISCO regions, which are important features to correctly evolve the accretion discs dynamics (see the spacetime analysis in \citealt{Combi2021}). While the Co24 metric formulation also comprehends the merger phase, we do not consider this feature in this work, with the exception of the Bondi-like test in which we evolved the binary until the merger moment (Sec \ref{sec:Bondi}).

\subsection{SKS implementation in \textsc{gizmo}}\label{sec:SKSImplementation}

The SKS formulation, as presented in Co24, has been coded in a public ready-to-use \texttt{C} implementation \citep{Combi2024SKS} that takes the BHs parameters as input data and returns the covariant metric from Eq.~(\ref{eq:SKSmetric}) at a given position. At each timestep, the needed inputs are the positions, velocities, spins, and masses of each BH. In our implementation, we took advantage of the BH particle structure already present in the Newtonian version of the code \citep[see][]{Franchini2022, Franchini2023, Franchini2024}, from which we exported the BH parameters. Hence, the binary is described by two BH particles that are evolved live, and their properties can be influenced by the interaction with the surrounding gas. Simulations presented in \citet{Franchini2023} showed the importance of evolving live binaries, highlighting crucial differences in the conservation of the total angular momentum of the system, in the binary separation evolution, and in the effects of gravitational torques.
In this work, however, we neglect the change in BH masses and spin driven by the gaseous disc for simplicity. This choice is reasonable as we are assuming our discs to have much smaller masses compared to the binary and therefore will have a negligible effect on its dynamical evolution. The focus here is instead to investigate the changes that the general relativity treatment will bring to the gas dynamics.

We integrate the equation of motion considering Post-Newtonian (PN) corrections up to the 2.5 PN order, i.e. including the contribution of the gravitational wave emission to the orbital evolution \citep{Blanchet2014}, to evaluate the BH trajectories. We refer to \citet{Franchini2024} for a complete presentation of the PN implementation in \textsc{gizmo}.\footnote{Note that the code can also work with pre-computed trajectories as those provided by the \textsc{cbwaves} tool \citep{CBwaves2012}.}

During the system evolution, gas can be accreted by the BHs. In our implementation, at each timestep, the code evaluates the distance of each gas particle to both BHs, and identifies the closest one. When such a distance becomes smaller than a user-defined excision radius,  the gas particle is removed from the simulation.
Since the superposed metric is written in the horizon penetrating KS coordinate system, we usually set the excision radius to reside inside the event horizon. As previously said, in this work, we do not change the BH properties when gas particles are accreted onto the horizons. This represents a good enough approximation if $M_{\rm BH}\gg M_{\rm gas}$, as we are assuming in our work.
In order to keep track of the accretion onto the BH during the simulation, at each timestep we also store the properties of the particles crossing the closest BH's event horizon radius.

We make use of the on-the-fly adaptive particle spitting approach, as presented in \citet{Franchini2022}. With this procedure, particles are split according to a scheme that depends on the particle distance to the domain centre. In practice, the mass resolution is doubled each time the gas crosses a set of defined radii, to better resolve the disc edge and the accretion streams inside the inner disc cavity. We refer to \citet{Franchini2022} for an in-depth presentation of this refinement approach. The only difference with the relativistic version is that, in the latter, while the two child particles inherit the hydrodynamical properties of the parent one, the metric coefficients are recomputed at the location of the two new particles.

Finally, when the resolution is poor, gas particles close to the BH's event horizon might extend the neighbor search to particles located on the other side of the BH. Such a case would either produce spurious results (as the gas cells should be causally disconnected from each other) or, in the worst case scenario, completely break the code. In order to avoid such events in our algorithm, the neighbor search in our implementation is limited to a maximum distance equal to the current distance of the particle from the closest BH. Note that we performed accretion tests varying the excision radius and with different neighbor search prescriptions, e.g. letting the kernel extend over the BH but preventing particle pairs whose line-of-sight is directly obscured by a BH from interacting. All the tests exhibited comparable results, without any statistically significant differences.

\section{Metric validation tests}\label{sec:Tests}

In this section, we present two GRHD tests we performed to confirm the correct implementation of the SKS metric in our code.\footnote{As an internal-consistency check, we also tested that the SKS metric recovers the single Kerr metric when one of the two BH masses was set to zero and the massive BH was set to rest by integrating circular orbits of collision-less particles.} We study here a set of inspiralling binaries at close and intermediate separations, surrounded by two different gaseous environments: a uniform plasma distribution and a thin circumbinary disc. We compare our results with those obtained using different evolution schemes. We focus on a comparison of the gas distribution and of the mass accretion rate onto the horizons.

\subsection{Bondi accretion}\label{sec:Bondi}
As a first test, we evolve an equal-mass binary initially placed on a circular orbit with separation of $12M$. For simplicity, in this test we do not include any live BH particle, but we define the BH location through pre-computed trajectories by \textsc{cbwaves}, which include PN corrections up to order 3.5. We embed the binary in an homogeneous gas distribution initially at rest extending up to $1500M$, and let the system evolve towards a Bondi-like solution around the binary system, similarly to the unmagnetized case studied by \citet{Cattorini2021}. Having such a large domain is important as we want the simulation to last for several orbits without including any particle injection at the outer boundary. At the same time, sampling the gas distribution with enough resolution to properly resolve the event horizon with several gas cells is crucial, as preliminary tests at low resolution showed spurious density fluctuations at the center of the domain when the number of cells was too low. 

Fulfilling both requirements at the same time would ideally require an extremely high number of particles, which would make the simulation unfeasible. To overcome this limitation, we decided to include adaptive particle splitting during the simulation, which operates between 50M and 1000M increasing the resolution by a factor of $2^{10}$ following a parabolic splitting function \citep[see][for details]{Franchini2022}. Moreover, we also mapped the initial conditions using the same splitting function employed during the simulation, resulting in 20 million particles with variable masses, instead of tens of billion particles required with a constant mass resolution everywhere. As we neglect the contribution of the gas self-gravity (see \citet{Franchini2021,Franchini2024b} for discs with self-gravity), the particle splitting does not introduce any spurious effect in the calculations while guaranteeing a very high resolution close to the binary already from the beginning. In order to prevent particles at the outer edge from leaving the domain, we further constrain the location and the hydrodynamic properties of the outmost 5 particle spherical shells to remain constant over time. 

\begin{figure}
    \centering
    \includegraphics[width=\columnwidth]{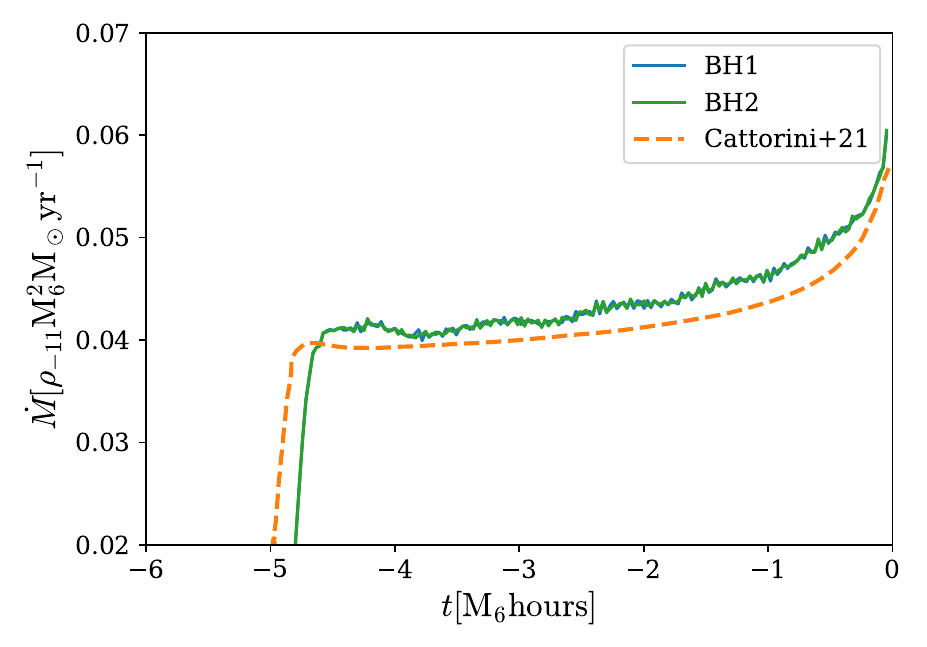}
    \caption{Mass accretion rate as a function of time in the Bondi-like test. The solution from \citet{Cattorini2021} is marked with an orange line.}
    \label{fig:mdotbondi}
\end{figure}

For this specific test, we evolved the system for the last 5 hours before merger, which occurs at $t=1772$ for a binary with $M=10^6\rm\, M_\odot$, assuming an ideal gas law with $\Gamma=4/3$.
In Fig.~\ref{fig:mdotbondi}, we show the accretion rate on each BH measured during the simulation, compared with the numerical relativity simulation results by \citet{Cattorini2021}. Given the symmetry of the system, the accretion rate is exactly the same on both BHs, as expected. After an initial rise, the accretion rate reaches a plateau which corresponds to the Bondi-like solution, in agreement with the results by \citet{Cattorini2021}. At later times, both simulations exhibit an increasing accretion rate as the two BHs approach merger. Compared to the numerical relativity results, the approximated metric slightly overestimates the gravitational pull onto the BHs, which results in a roughly 10 percent higher accretion rate \citep{Combi2024}. Considering i) the different space-time evolution in the two cases, ii) differences in the numerical techniques employed and the discretization scheme, and also iii) the maximum resolution close to the horizon, which in our case is about 0.4$M$, much lower than the resolution reached by  \citet{Cattorini2021}, i.e. $0.02M$, these differences do not look particularly significant, especially when we consider that our simulation required only 40 hours on 48 CPUs. 
Note also that in  numerical relativity simulations, the space-time perturbations due to the two BHs close to merger are expected to differ from the analytic metric by Co24, potentially causing  inaccuracies also in the accretion rate calculation. Moreover, the accretion rates in grid-based simulations are typically estimated by integrating the mass flux at the horizon surface, whereas in \textsc{gizmo} we can directly tag the particles crossing it. 
We can therefore conclude that our treatment of the binary dynamics is sufficiently robust and we can proceed by running simulations of binaries embedded in circumbinary discs.

\subsection{Circumbinary discs evolution}\label{sec:CBD}

We performed a direct comparison between a circumbinary disc evolved with the Newtonian version of \textsc{gizmo} and one evolved with our general relativistic (GR) implementation in order to study the effects of the SKS metric implementation on the gas evolution.

For both runs, we considered the same initial conditions, which correspond to the ones used in \citet{Franchini2022, Franchini2023} and produced with the code \texttt{PHANTOM} \citep{Price2017}. We initialized $10^6$ particles sampling a Newtonian accretion disc extending from $R_{\rm in}=150 M$ to $R_{\rm out}=1000 M$. The initial $H/R$ ratio was $0.1$. The BH binary has been initialized on a circular orbit with a separation $d=100M$ and the BHs are non-spinning.
We note that the initial configuration was fully Newtonian. This assumption is well justified because the initial binary separation was sufficiently large and the disc was far from the BHs to ensure that relativistic corrections played only a minor role.
As an equation of state, we used the ideal gas relation, with $\Gamma=5/3$, both in the initial conditions and during the evolution.

We performed the first run using the Newtonian version of the code, where the gas is evolved according to the classical hydrodynamic equation and the binary potential is given by the superposition of the two Keplerian potentials generated by the BH particles.
The second run was evolved with the general relativistic scheme, where the gas follows the equations of general relativistic hydrodynamics as presented in \citet{Lupi2023} and with the improvements described in \citet{Fedrigo2025}, evaluated on the SKS metric.
In both cases, we employed the MFM method, adaptive particle splitting \citep{Franchini2022} and the BH dynamics was evaluated considering PN correction terms up to the $2.5$ order. For simplicity, we decided to evolve both discs as inviscid, thus without either viscosity or magnetic field prescriptions (the latter will be the focus of an upcoming work). Although idealised, this test allows us to focus on the relativistic effects of the binary metric on the gas only, without any additional processes which might alter the comparison. The angular momentum transport therefore occurs only via asymmetries in the flow and possible numerical dissipation.\\ A more comprehensive description of the physics within accretion discs would require solving the radiative transport equations during the hydrodynamical evolution of the system, possibly affecting the thermodynamics and the structure of the disc \citep{Kaaz2024}. However, such a prescription would be computationally really heavy. A much simpler approach would be to artificially cool the gas to a target entropy, as done in \citet{Noble2012} or \citet{Combi2022}, where the evolved discs appeared thinner with less vertical pressure support \citep{dAscoli2018}. In our simulations, however, we decided to neglect any gas cooling prescription to keep the problem as simple as possible. As a consequence, we expect our discs to become warmer and thicker over time.
We let the two systems evolve until $t=200 T_0$, where $T_0$ is the orbital period of the binary at $t=0$.

\begin{figure*}
\begin{center} 
\includegraphics[width=.39\textwidth, trim=0 0 4.5cm 0,clip]{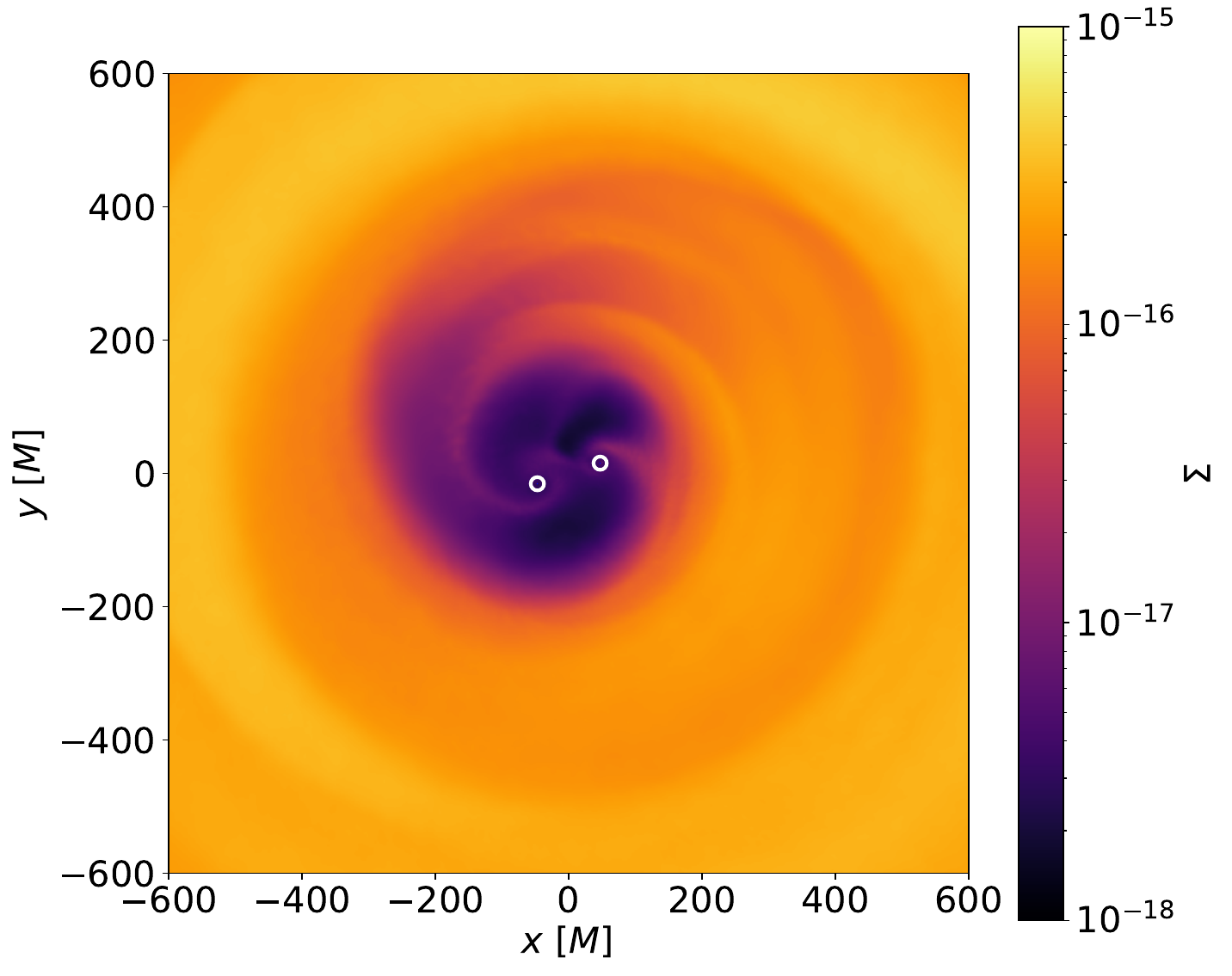}
\includegraphics[width=.49\textwidth]{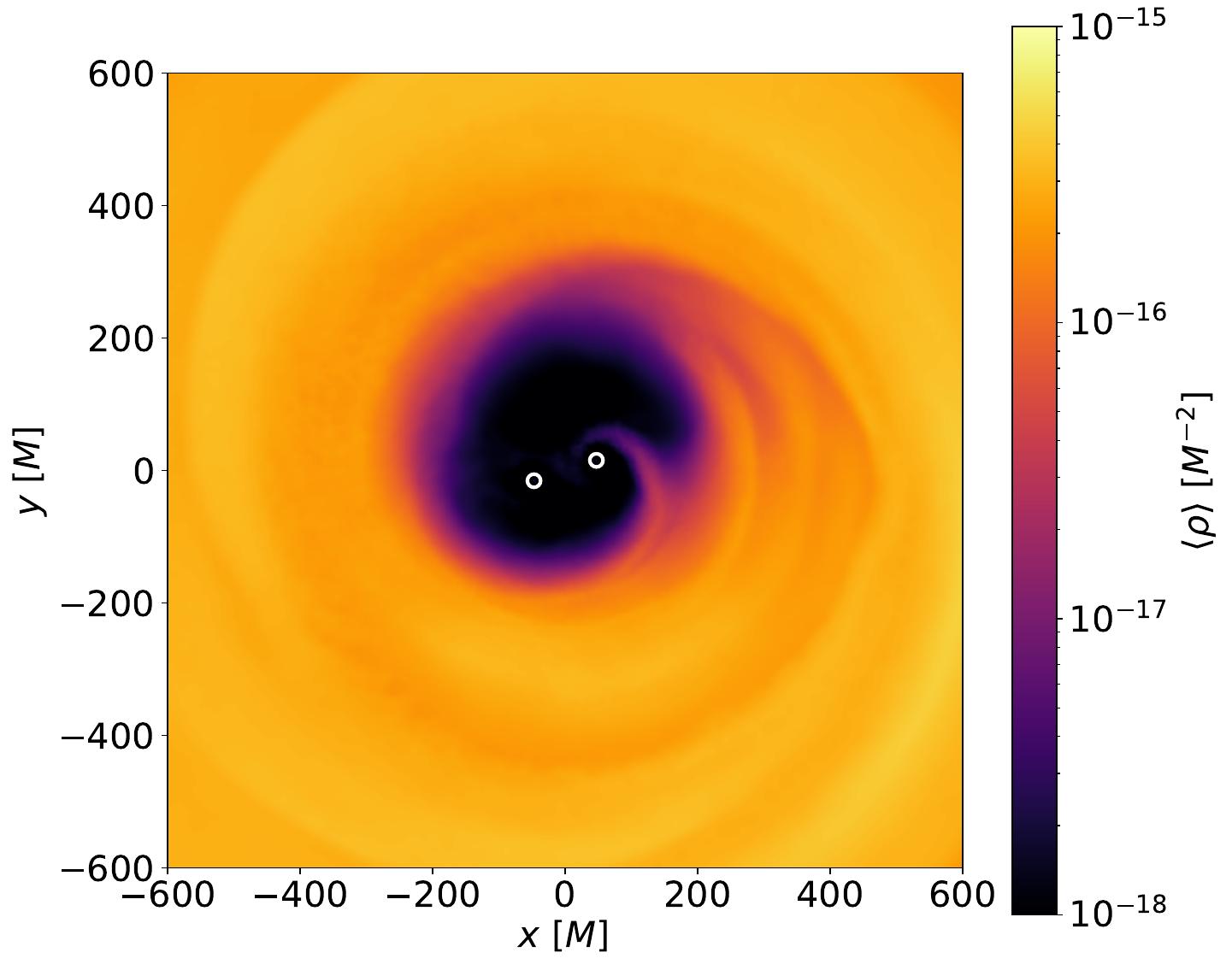}
\end{center}
\caption{Density map of the central part of the circumbinary disk for the general relativistic run (left) and the Newtonian run (right), at $t=200T_0$. Density is volume-averaged along the z direction. The white circles mark the BHs' event horizons, scaled up by 10 times for visualisation purposes.}\label{fig:CBDDensityMaps}
\end{figure*}

In Fig.~\ref{fig:CBDDensityMaps}, we show density maps of the two evolved circumbinary discs at $t=200T_0$.
In the Newtonian snapshot (right-hand panel), an empty, slightly elliptical cavity is formed around the binary, carved by the torques produced by the orbiting BHs. The edge of the cavity is found at approximately $r\approx200- 250 M$ i.e. $\sim 2d$, consistent with previous simulations. A faint gas stream can be seen accreting from the disc to one BH's horizon.
\\On the other hand, the general relativistic (left-hand panel) evolution displays a slightly smaller and more circular internal cavity and lower density in the inner region of the disc. The inner cavity extends up to $r\approx 150-200M$, while a second under-dense region can be seen extending up to $r\approx 500M$. As expected, both discs get thicker over time, reaching $H/R$ ratios of $\sim 0.15$.

At late times, after $\approx100$ orbital periods, both discs appear asymmetric, with a $m=1$ weak overdensity, usually refereed to as "lump" \citep{Shi2012, Noble2012, MignonRisse2023}. This feature is believed to be caused by streams of gas falling in the central cavity and redirected outward toward the inner edge of the circumbinary disc, and it is usually observed in both magnetised and non-magnetised circumbinary discs \citep{Noble2021} around either equal-mass or unequal-mass binaries, provided that the mass ratio is high enough. The gas accumulates in a region that is usually a few times denser than the rest of the inner disc. The origin of the lump is often identified with the Rossby-Wave Instability (RWI), which might occur in the innermost regions of the circumbinary disc \citep{MignonRisse2023}. The RWI is expected to develop in the presence of local maxima in the fluid vortensity \citep{Lovelace1999}. A simpler criterion for the development of the RWI requires the minima of epicyclic frequency $\kappa$ to be $\min(\kappa^2)\lesssim 0.5$-$0.6 \Omega^2$ \citep{ChangYoudin2024}. This condition can be satisfied in the presence of strong density gradients, such as the one between the cavity and the inner edge of the circumbinary disc.

We computed the ratio $\kappa^2/\Omega^2$ within our circumbinary disc, finding values $\sim 0.5$ in the innermost regions with smaller values limited to a very narrow radial range and only in our GR run. Our density gradient at the cavity edge does not seem strong enough to develop RWI, i.e. we find a "lump" with density contrast well within a factor of two. The reason for such a weak "lump" likely lies in the choice of the equation of state, i.e. adiabatic without cooling, and the dimensionality of the problem. For instance, prominent lumps have been found in circumbinary disc simulations employing an isothermal fluid \citep{Shi2012} or cooling prescriptions that efficiently dissipates extra heating from shocks \citep{Noble2012}. In the adiabatic case, instead, a prominent lump is found in two-dimensional simulations which prevent the gas from re-distributing along the z-direction, boosting density gradients and, as a consequence, more easily triggering the RWI \citep{MignonRisse2023}.

\begin{figure}
\begin{center} 
\includegraphics[width=0.48\textwidth]{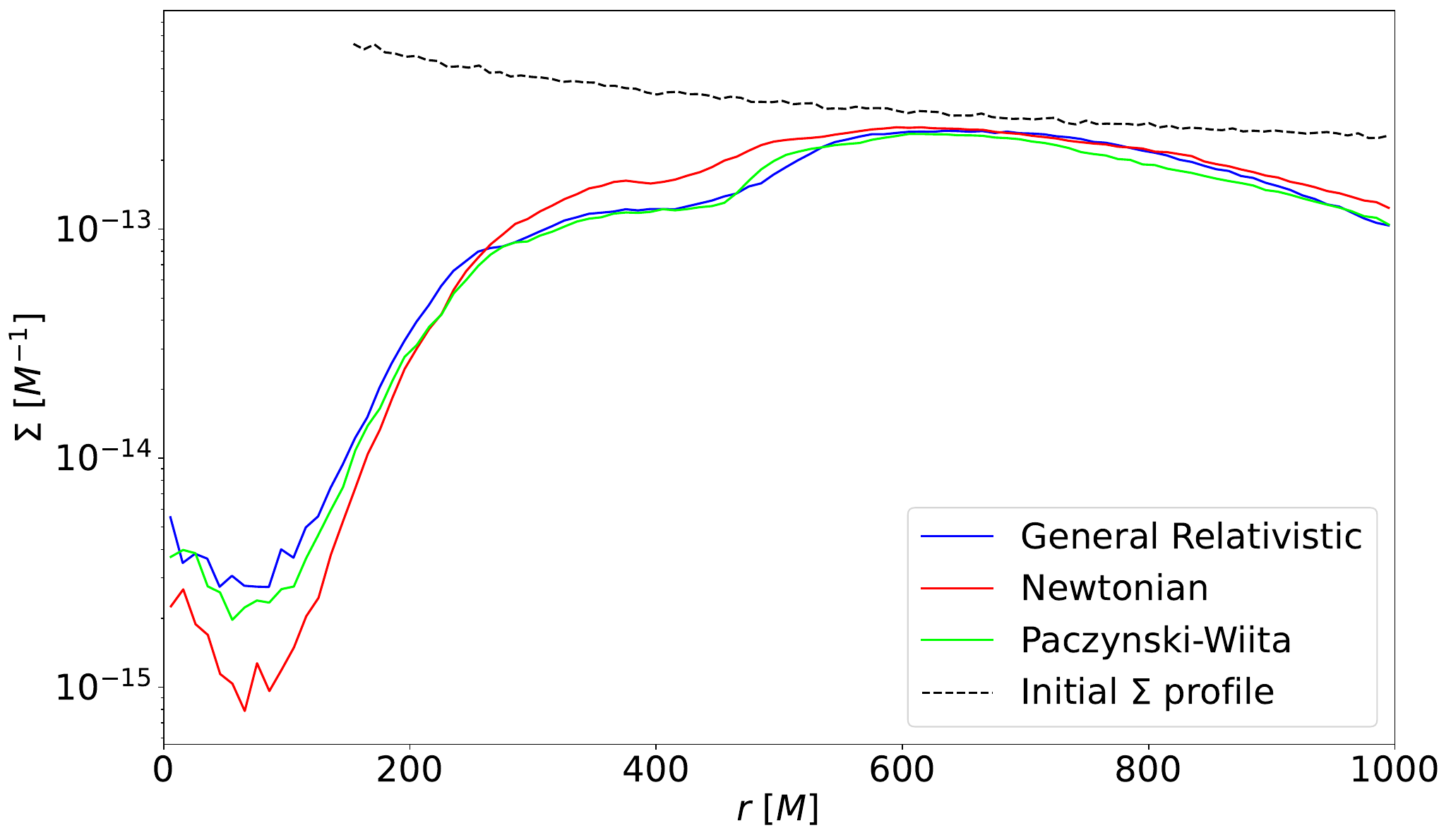}
\end{center}
\caption{Surface density $\Sigma$ radial profile for the GR (blue), for the Newtonian (red), and for the Paczynski-Wiita (lime) runs at $t=200T_0$. The initial $\Sigma$ profile is marked with a black dashed line.}\label{fig:CBDSurfaceDensity}
\end{figure}

For a more quantitative analysis, in Fig.~\ref{fig:CBDSurfaceDensity} we plot the surface density $\Sigma$ radial profile for both runs at $t=200T_0$. In the innermost region, the Newtonian evolution yields a surface density $2-3$ times lower than the GR case, indicating that less gas is entering the central cavity. Between $r\approx 250M$ and $r\approx 500M$, a small drop in the $\Sigma$ profile highlights the under-dense region discussed above. This feature is more pronounced in the relativistic run, with a drop by a factor of $2$. We elaborate on this discrepancy further below in this section.

Finally, we evaluated the total mass accretion rate onto the BHs' event horizons $\dot{M}$ and show it in Fig.~\ref{fig:CBDMdot}. As expected, after an initial transient (highlighted by a shaded grey region in the plot), the Newtonian binary (red line) exhibits low gas accretion, with a slow decay over time. We recall that there is no viscosity or magnetic field prescription in the simulations, which are usually the principal processes invoked to transport angular momentum and promote accretion. We note that the adaptive particle splitting procedure may introduce a small diffusivity in the evolution scheme at the radii where the resolution is refined.\footnote{During testing, we found that the splitting procedure in the GR scheme is slightly more diffusive than the Newtonian one. However, in our runs this difference results in a $\dot{M}$ increase of $10^{-2}$ $\langle\dot{M}\rangle$.} On the other hand, its adoption in our work allows us to better resolve the gas morphology within the cavity \citep{Franchini2022, Duffell2024}.
We opted to use the adaptive resolution to correctly resolve the internal structure at the cost of the introduction of a small diffusivity.

The GR binary (blue line in Fig. \ref{fig:CBDMdot}) shows a non-zero accretion rate characterized by a constant average trend and a quasi-periodic variability related to the inner disc orbital motion. These kinds of oscillations are also seen and studied in depth in many other works \citep{Paschalidis2021,Combi2022,Cattorini2022,Bright2023,Fedrigo2024,Manikantan2025}.
We note that the presence of a non-zero accretion rate in the GR run is consistent with the density drop found in the $\Sigma$ profile between $r \approx 200-500 M$ (Fig. \ref{fig:CBDSurfaceDensity}). The accreting gas comes from this inner part of the disc, which gets less dense over time.\footnote{We let the relativistic run evolve until $t=600T_0$; a bigger cavity with radius $r\simeq 500 M$ is carved around the BHs}

Given the differences between the Newtonian and GR numerical schemes, such a difference might arise due to multiple reasons, either numerical or physical. For this reason, we performed tests using more diffusive slope limiters in the Newtonian binary evolution, which resulted in a slightly higher mass accretion rate compared to the reference case. However, even with a higher numerical diffusion, the Newtonian accretion rate remains lower than the GR run, suggesting that the diffusivity of the scheme cannot explain the different inflow on the binary in the two cases.\footnote{We also tested whether the initial condition had any effect on the late-time dynamics of the gas in the disc, finding none.}

\subsubsection{Post-Newtonian validation test}
\begin{figure}
\begin{center} 
\includegraphics[width=0.48\textwidth]{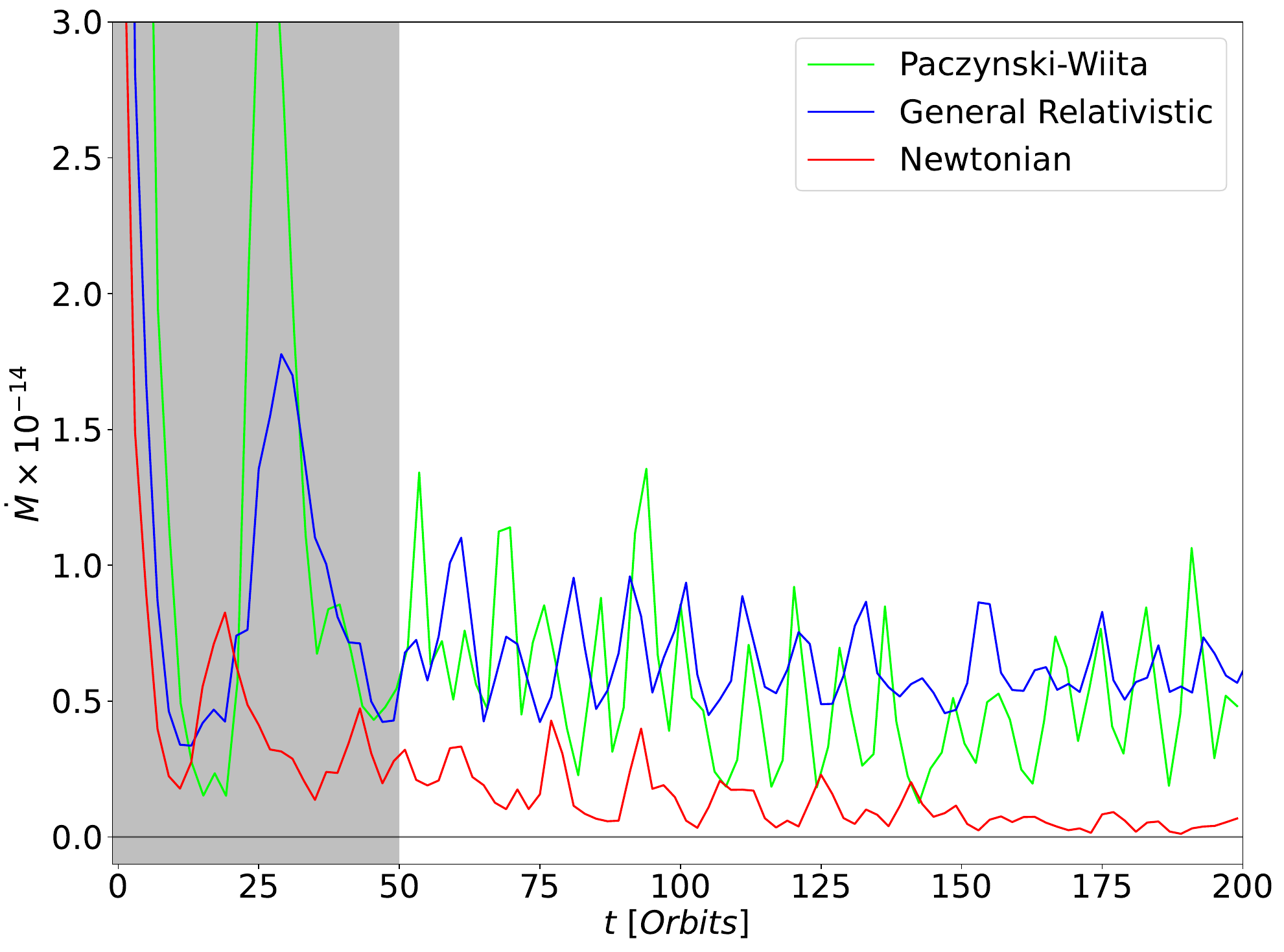}
\end{center}
\caption{Mass accretion rate as a function of time for the GR (blue), the Newtonian (red), and the Paczynski-Wiita (lime) binaries. The transient phase, caused by the initial data relaxation, is marked with a grey shaded region.}\label{fig:CBDMdot}
\end{figure}

A physical explanation would be instead that the higher gas accretion rate is due to the non-Keplerian nature of the potential. To test this hypothesis, we performed a run using the Paczynski-Wiita (PW) potential implemented in the Newtonian scheme. The PW potential for a binary can be written as
\begin{equation}\label{eq:PWpotential}
    \Phi(\textbf{x}) = -\frac{GM_1}{(|\textbf{x}-\textbf{x}_1|-\textbf{r}^{EH}_{1})} -\frac{GM_2}{(|\textbf{x}-\textbf{x}_2|-\textbf{r}^{EH}_{2})}
\end{equation}
where $|\textbf{x}-\textbf{x}_i|$ is the distance between the point where the potential is evaluated and the i-th BH and $\textbf{r}_i^{EH}$ is the radius of the i-th BH's event horizon. This Post-Newtonian potential \citep{PaczynkiWiita1980} only produces $2/3$ of the apsidal precession produced by a relativistic Schwarzschild potential (in the single central object scenario), but it is extremely easy to implement and fast to compute. For this test, we started from the same initial conditions employed in the other two runs, and let the system evolve for $\sim200$ orbits. The resulting $\Sigma$ profile is shown by the lime line in Fig.~\ref{fig:CBDSurfaceDensity}, whereas the $\dot{M}$ is reported, again with a lime line, in Fig.~\ref{fig:CBDMdot}. We clearly see that the surface density profile in the Post-Newtonian run is very similar to the GR case. After the usual initial transient, the mass accretion rate displays a trend comparable with the GR case, with stronger oscillation but similar averaged values. In fact, the average $\dot{M}$ computed between $50$ and $200$ orbits is $\langle\dot{M}\rangle = 5.4\cdot10^{-15}$ for the PW run, while it amounts to $6.4\cdot10^{-15}$ and $1.2\cdot10^{-15}$ in the GR and Newtonian runs, respectively.

\begin{figure}
\begin{center}
\includegraphics[width=.4\textwidth,trim=1cm 0 1cm 0]{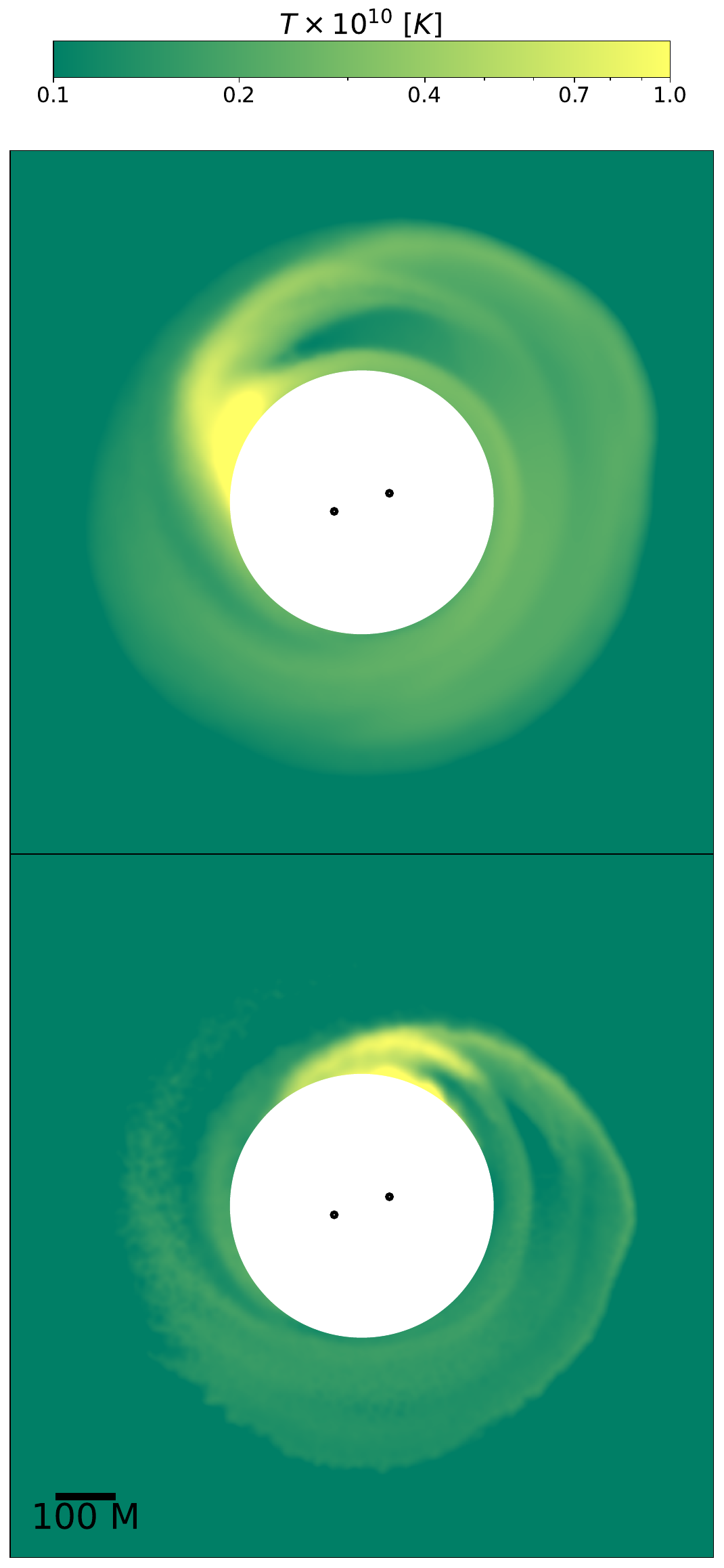}
\end{center}
\caption{Density-weighted gas temperature distribution at $t=200T_0$ for the GR (top panel) and the Newtonian (bottom panel) run. The inner cavity ($r\lesssim 225M$) is masked with a white patch. The temperature is normalized for a binary of $M_{b}=10^6 M_\odot$.}\label{fig:T}
\end{figure}

We suggest that relativistic apsidal precession of the gas orbiting a BH binary may play a significant role in the circumbinary disc scenario \citep{Henon1965,Paczynski1977,RudakPaczynski1981}. When the trajectories of the orbiting gas are forced to cross by apsidal precession, shocks are generated. During shocks, part of the kinetic energy is converted into thermal energy, and  angular momentum is transferred outward, resulting in a net gas inflow towards the centre. To validate this idea, we show the density-weighted gas temperature distribution at $t=200T_0$ in Fig. \ref{fig:T}, obtained by measuring the pressure and density from the simulation and then using $P=\rho k_bT/m_p\mu$. As expected, the GR disc tends to be about a factor of 2-3 hotter than its Newtonian equivalent in the region affected by shocks, in agreement with our hypothesis. We note, also, that due to relativistic apsidal precession, the hottest area found at the inner edge of the relativistic disc is shifted forward in the direction of the disc rotation relative to the Newtonian case. We note that a gas cooling prescription could have mitigated this heating process in both our runs.

The presence of apsidal precession can therefore justify the higher mass accretion rate found in our GR and PW discs runs when compared with the Newtonian simulation. Even though this effect produces only a small perturbation to the gas trajectory, it can be enough to trigger a faint, but non-negligible accretion process. Note that, when a more realistic scenario is considered, e.g, including the presence of an effective viscosity or of magnetic fields triggering MRI, the inflow rate we observed will likely be subdominant, thus going unnoticed in the presence of more efficient accretion mechanisms.

\subsubsection{Angular momentum transport and effective viscosity}

In order to measure the angular momentum transport within the disc, we computed the effective viscosity $\alpha_{\rm eff}$ in our discs in different ways. 
First, we assumed the prescription by \citet{ss1973} who modelled the viscous stress tensor as simply proportional to the pressure in the disc, $\alpha$ being the constant of proportionality. From this follows the expression for kinematic viscosity $\nu\equiv \alpha_{\rm eff}c_{\rm s} H$. For a disc in steady state the kinematic viscosity is related to the mass accretion rate onto the central object and the disc surface density as $\nu=\dot{M}/3\pi\Sigma$. 
We compute the accretion rate radial profile within the disc by measuring the inward flux of material through cylindrical surfaces at different radii.
We averaged the results over the last 10 binary orbits, and we obtained values of $\alpha_{\rm eff}$ ranging between $10^{-3}$ and $10^{-2}$. The GR run exhibits values $3-5$ higher than the Newtonian one. We note that this quantity estimates the global angular momentum transport, taking into account both the contribution from hydrodynamical stresses and from numerical dissipation.

Secondly, we computed the effective viscosity using its definition $\langle R^r_{\ \phi}\rangle=\alpha_{\rm eff}\langle P\rangle$, where $\langle R^r_{\ \phi}\rangle=\langle\rho h \ \delta u^r \ \delta u_\phi\rangle$ is the Reynolds stress tensor, $\delta u^\mu = u^\mu-\langle\rho u^\mu\rangle/\langle \rho \rangle$ is the velocity fluctuation relative to the mean motion and $\langle \cdot \rangle$ is a time and azimuthal average operator. We use the relativistic formalism presented in \citet{Noble2012}, and we averaged again over the last 10 binary orbits. Using this prescription, we obtained effective viscosity values up to $5\times 10^{-3}$ and $2\times10^{-3}$ in the outermost part of both discs for the GR and Newtonian run, respectively. A somewhat irregular behaviour is instead seen in the inner part of the disc, between $r\approx2.5d$ and $r\approx5d$, where the development of spiral waves and asymmetries in the gas distribution results in a noisier local effective viscosity, with values reaching $10^{-2}$ \citep{MignonRisse2025}. At the cavity edge, $\alpha_{\rm eff}$ reaches a maximum value of $\sim 2.5\times10^{-2}$.
As suggested in \citet{Noble2012}, we also computed the Reynolds stress tensor as $\langle R^r_{\ \phi}\rangle = \langle T^r_{\ \phi}\rangle - \langle A^r_{\ \phi}\rangle$, where $T$ is the stress-energy tensor, $\langle A^r_{\ \phi}\rangle = \langle \rho l\rangle\langle\rho h u^r\rangle/\langle\rho\rangle$ and $l=-u_\phi/u_t$ is specific angular momentum. As expected, this equivalent formulation yielded results compatible with the $\alpha_{\rm eff}$ estimation via the fluid velocity fluctuations.

Comparing the different $\alpha_{\rm eff}$ estimations, we can conclude that the angular momentum transport in our discs is dominated by the hydrodynamic stresses from fluid asymmetries. However, this effect yields an effective viscosity $1-2$ orders of magnitude lower than values usually adopted in viscous disc simulations, e.g. $\alpha \sim 0.1$ \citep{Franchini2024}.

\subsubsection{Minidiscs morphology and dynamics}

\begin{figure}
\begin{center} 
\includegraphics[width=.45\textwidth,clip]{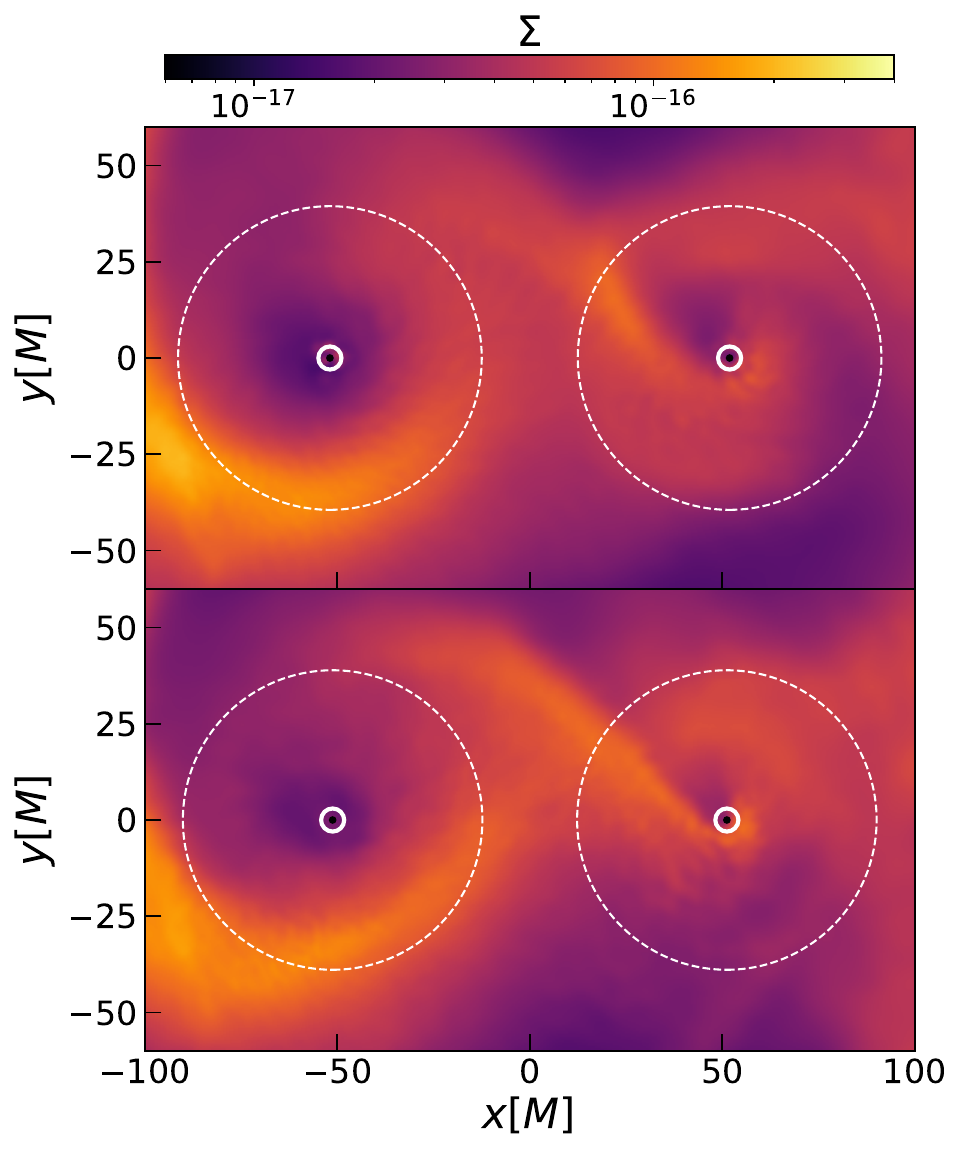}
\end{center}
\caption{Density maps of the minidiscs during a filling phase (after $\simeq 60$ BH orbits from the start of the run) for the GR (top panel) and Newtonian (bottom panel) run. The black dots mark the event horizons, the solid lines the ISCO radii and the dashed lines the Roche lobe radii. Snapshots are aligned with the binary axis.}\label{fig:Minidiscs}
\end{figure}

As a final analysis, we consider here the morphology of the accreting structures that form close to the BHs' horizons, usually called ``minidiscs'', which are expected to be affected by relativistic corrections.
To define these structures, we considered the gas present within each BH's Roche lobe \citep{Eggleton1983}. In the case of an equal mass binary, this region can be approximated as a sphere centred on the BH, with an equivalent-volume radius of $R_{eq}\simeq0.38d$ \citep{Leahy2015}. We note that the very low angular momentum transport and the consequent low density in the central cavity lead to transient and very variable "minidisc" structures (Fig. \ref{fig:Minidiscs}), hardly comparable in morphology to the ones found in simulations with stronger angular momentum transport, e.g. \citep[GRMHD]{Noble2012,Combi2022} or \citep[viscous]{Franchini2022}. However, they serve our purpose to study general relativistic effects caused by the metric.

By studying the time evolution of minidiscs in the two runs, we recognise a more uniform and homogeneous gas distribution in the relativistic case, while the Newtonian one tends to be more asymmetric. Furthermore, the latter is more prone to strong density fluctuations, indicating moments where the Roche lobes are almost empty.
The more stable evolution and the more homogeneous gas distribution of the GR minidiscs can be understood in terms of the higher accretion rate in the cavity. \\Relativistic apsidal precession plays an important role here too. 
In our simulation, the gas streams falling toward the relativistic BHs circularise, producing more homogeneous gas structures around the BHs and effectively mitigating density asymmetries in the minidiscs.

\begin{figure}
\begin{center} 
\includegraphics[width=.5\textwidth, trim=1cm 0 0 0,clip]{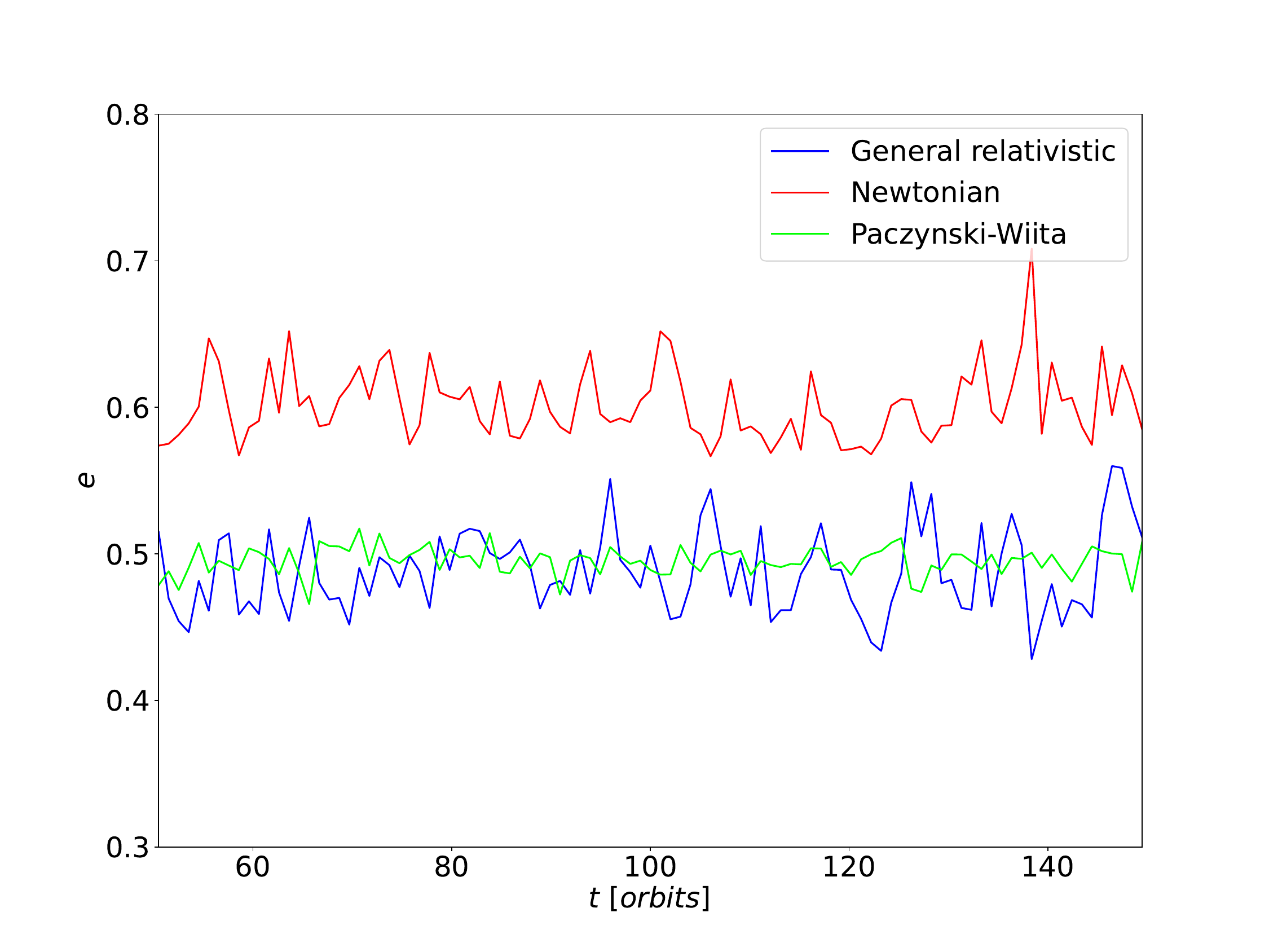}
\end{center}
\caption{Average gas eccentricity within the minidiscs as a function of time for the GR (blue), the Newtonian (red), and the Paczynski-Wiita (lime) runs.}\label{fig:MinidiscsEccentricity}
\end{figure}

We evaluated the minidiscs eccentricity in our simulations from the gas energy within the Roche lobes.
We performed this analysis between $t=50T_0$ and $t=150T_0$, i.e. excluding the initial transient and the final part of the simulations in which the Newtonian cavity is the emptiest, and we plot the results in Fig. \ref{fig:MinidiscsEccentricity}.
While the Newtonian minidiscs show an eccentricity that oscillates around $e\approx 0.6$, the GR and PW ones average approximately to $e\approx 0.5$. Hence, the relativistic minidiscs tend to be slightly more circular than the Newtonian ones. We note that the circulatization effect is small because we are considering non-spinning BHs, whose apsidal precession effect is less strong if compared to spinning BHs.

Finally, we note that, while the PW prescription can produce similar results to our GR formulation in the scenario we considered here, it cannot emulate the spin of the BH and its effects on the gas dynamics. Also, the PW potential is divergent at the event horizon, which can lead to numerical instabilities and artifacts.

To conclude, our results highlight the importance of considering general relativistic effects in the evolution of circumbinary discs and BHs' minidiscs, as they play an important role even at distances as large as hundreds of gravitational radii and thousands of orbits before the binary merger.

\section{Conclusions}\label{Sec:Conclusion}

In this paper, we presented our implementation of the SKS metric from Co24 in the relativistic scheme of the \textsc{gizmo} code. Now the code can simulate BH binaries at intermediate separations ($10^1M \lesssim d\lesssim10^3M$) with great computational efficiency and accuracy, taking into account relativistic effects in the gas and magnetic field evolution.

We tested our implementation by comparing our results with a numerical relativity simulation of a BH binary embedded in a uniform gas distribution, obtaining good agreement in the gas evolution behavior and in the estimated mass accretion rate onto the BH horizons.
Then, we compared the evolution of two circumbinary discs evolved using the Newtonian and relativistic schemes in \textsc{gizmo}. Both simulations were stable and produced reasonable results. The GR run, however, exhibited a higher gas accretion rate onto the horizons and a lower density region in the innermost part of the disc, between $r\approx 2.5d$ and $r\approx5d$. We found that the presence of a non-Keplerian potential is the cause of this phenomenon, as demonstrated by a Post-Newtonian simulation performed with the standard version of \textsc{gizmo}.

With this work, we highlight the importance of including relativistic effects in simulations of circumbinary discs around BH binaries approaching merger, even at scales not yet accessible with numerical relativity simulations.

In a future work, we plan to explore the parameter space of long-evolved magnetised circumbinary discs, searching for characteristic features that could suggest EM signatures.

\begin{acknowledgements}
We thank the anonymous referee for very useful comments that improved the quality of the manuscript. We thank Daniel Price and Enrico Ragusa for useful discussions.
AL acknowledges support from PRIN MUR “2022935STW".
GF thanks Sormani Mattia Carlo and Zolt\'{a}n Haiman for their fruitful suggestions on the angular momentum transport interpretation. MB acknowledges support from the Italian Ministry for Universities and Research (MUR) program “Dipartimenti di Eccellenza 2023-2027”, within the framework of the activities of the Centro Bicocca di Cosmologia Quantitativa (BiCoQ). AF acknowledges financial support from the Unione europea - Next Generation EU, Missione 4 Componente 1 CUP G43C24002290001. 
\end{acknowledgements}

\bibliographystyle{aa}

\bibliography{bibliografia}
\end{document}